\documentclass{ifacconf}

\usepackage{graphicx,subcaption}      % include this line if your document contains figures
\graphicspath{{Figures/}}

\usepackage{amsmath}
\usepackage{amssymb}
\usepackage{helvet}
\usepackage[acronym, toc]{glossaries}
\usepackage{siunitx}
\usepackage{tikz,bm,color}
\usepackage{enumerate}
\usepackage{mathtools, cuted}
\usepackage{pgfplots}
\usepackage{pgfplotstable}
\usetikzlibrary{external,calc,patterns,decorations.pathmorphing,decorations.markings,decorations.text,arrows,shapes,positioning, backgrounds, intersections}
\usetikzlibrary{spy}
\usepgfplotslibrary{fillbetween}
\usepackage{caption}
\usepackage{hhline}
\usepackage{ctable}
\usepackage{pbox}

\definecolor{cadmiumgreen}{rgb}{0.0, 0.42, 0.24}
\definecolor{PIcolor}{rgb}{0, 0, 1.0}
\definecolor{STSMCcolor}{rgb}{0.929, 0.427, 0.067}
\definecolor{NACcolor}{rgb}{0.0, 0.42, 0.24}
\definecolor{InIcolor}{rgb}{0.247, 0.475, 0.749}
\definecolor{limitColor}{rgb}{0.247, 0, 0}
\pgfplotsset{compat=newest}
\usepackage{balance}

\newtheorem{assumption}{Assumption}

\newtheorem{remark}{Remark}

\usepackage{algorithm}
\usepackage{algpseudocode}
\usepackage[normalem]{ulem}
\usepackage{slashbox}
\newcommand\T{\rule{0pt}{2.6ex}}       % Top strut
\newcommand\B{\rule[-1.2ex]{0pt}{0pt}} % Bottom strut

\newacronym{AMB}{AMB}{Active Magnetic Bearings}
\newacronym{PMB}{PMB}{Passive Magnetic Bearings}
\newacronym{PMSM}{PMSM}{Permanent Magnet Synchronous Motor}
\newacronym{PMSMs}{PMSMs}{Permanent Magnet Synchronous Motors}
\newacronym{SMC}{SMC}{Sliding Mode Control}
\newacronym{ISS}{ISS}{Input-to-State Stable}
\newacronym{SISO}{SISO}{Single-Input Single-Output}
\newacronym{PID}{PID}{Proportional-Integral-Differential}
\newacronym{PI}{PI}{Proportional-Integral}
\newacronym{P}{P}{Proportional}
\newacronym{STSMC}{STSMC}{Super-twisting Sliding Mode controller}
\newacronym{PE}{PE}{Persistence of Excitation}
\newacronym{ULES}{ULES}{Uniformly Locally Exponentially Stable}
\newacronym{GES}{GES}{Globally Exponentially Stable}
\newacronym{UB}{UB}{Uniformly Bounded}
\newacronym{UGB}{UGB}{Uniformly Globally Bounded}
\newacronym{UGAS}{UGAS}{Uniformly Globally Asymptotically Stable}
\newacronym{UAS}{UAS}{Uniformly Asymptotically Stable}
\newacronym{MAE}{MAE}{Maximum Absolute Error}
\newacronym{RMSE}{RMSE}{Root Mean Square Error}
\newacronym{VSC}{VSC}{Variable-Structure Control}

\makenoidxglossaries
\usepackage{natbib}        % required for bibliography
\begin{document}
\begin{frontmatter}

\title{Sliding Mode Control of Active Magnetic Bearings - A Cascaded Architecture} 

%\thanks[footnoteinfo]{Sponsor and financial support acknowledgment
%goes here. Paper titles should be written in uppercase and lowercase
%letters, not all uppercase.}

\author[First]{Dimitrios Papageorgiou} 
\author[Second]{Ilmar Santos}

\address[First]{Technical University of Denmark, Department of Electrical and Photonics Engineering, Elektrovej 326, 2800 Kgs Lyngby, Denmark (e-mail: dimpa@dtu.dk).}
\address[Second]{Technical University of Denmark, Department of Civil and Mechanical Engineering, Koppels All{\'e} 404, 2800 Kgs Lyngby, Denmark (e-mail: ilsa@dtu.dk).}

\begin{abstract}                % Abstract of not more than 250 words.
	Accurate and robust positioning of rotor axle is essential for efficient and safe operation of high-speed rotational machines with active magnetic bearings. This study presents a cascaded nonlinear control strategy for vertical axial positioning of an active magnetic bearing system. The proposed scheme employs two sliding mode controllers for regulating rotor vertical position and current and an adaptive estimator to invert the nonlinear input mapping. Uniform asymptotic stability is proven for the closed-loop system and the efficacy and performance of the proposed design is evaluated in simulation.
\end{abstract}


\end{frontmatter}
%===============================================================================

\section{Introduction}
	\gls{AMB} are becoming more and more popular in rotating machine systems. This is due to their feature of facilitating contactless rotation, hence rotary machines with \gls{AMB} do not require lubricants, exhibit less wear and support high rotational speed limits. A wide integration of \gls{AMB} into rotary equipment can revolutionise the operating conditions in energy production systems such as flywheels.

Active control of \gls{AMB} systems is instrumental for accurate positioning of the rotor axle, which is necessary for safe and efficient operation of high-speed rotational machines. Feedback control strategies in particular, can ensure robust operation of high-speed rotational machines so that they become available in industrial motion systems, allowing energy-efficient technology to enter manufacturing and production.

Active control of \gls{AMB} systems has been extensively studied in the literature. Robust linear control approaches such as $H_{\infty}$ loop shaping were investigated for disturbance rejection \citep{schroder1997a} and in combination with closed-loop identification \citep{junfeng2011a}. $H_{\infty}$ control of uncertain \gls{AMB} systems was studied in \citep{lee2013a} using a linear parameter-varying systems framework, while the authors in \citep{fittro2002a} proposed a $\mu$-synthesis approach. Adaptive control methods were investigated in several studies such as \citep{gibson2002a} and \citep{long1996a}, where the authors proposed an adaptive backstepping control scheme for compensating for uncertain load change and unbalance disturbance. Adaptive backstepping control was employed in \citep{dong2013a} for a linearised \gls{AMB} system with small parametric uncertainties. Nonlinear backstepping control was proposed in \citep{motee2002a} for regulation of the radial motion of an \gls{AMB} system. An adaptive extension of this scheme was studied in \citep{xu2022a}, where only the mechanical dynamics were considered. \gls{SMC} approaches for linearised \gls{AMB} plants were proposed in \citep{huynh2016a}, \citep{kang2010a} and \citep{jang2005a}. \gls{SMC} for radial motion regulation was developed in \citep{wang2016a}. The nonlinear dynamics of \gls{AMB} systems was considered in several control approaches, however with linear approximations of the magnetic forces such as in \citep{saha2020a} and \citep{kandil2018a}.

The majority of the reported work on control of \gls{AMB} employ linearisation of the magnetic forces expressions, which limits the operating range of the model. Static input inversion was used in \citep{mouille1992a} to compute the appropriate voltage that would generate the demanded control forces. A similar approach was adopted in \citep{song1996a}, where nonlinear adaptive control was designed for an \gls{AMB} system. The inversion of the nonlinear input mapping was achieved by solving a set of second-order algebraic equations at every time instant. Such approaches can be problematic when there is ambiguity regarding multiple solutions, eventually leading to non-smooth current or voltage reference signals.

This paper proposes a cascaded nonlinear control strategy for the vertical axial positioning of the rotor shaft in an \gls{AMB} system. The control scheme comprises nonlinear control laws based on sliding mode principles for the mechanical and current dynamics that are designed separately. An adaptive estimator is designed for inverting the nonlinear input mapping in the \gls{AMB} system. The inherent modularity of this architecture allows not only for easier analysis but also for integration of different methods in the control cascade. More specifically, the contributions of this study pertain to the following:
\begin{itemize}
	\item Development of a modular cascaded control architecture, into which alternative control and estimation algorithms can be integrated.
	\item Development of a \emph{dynamic} estimation strategy for inverting the nonlinear input mapping between desired force and current.
	\item Stability analysis of the cascaded closed-loop system.
\end{itemize}

The remainder of the paper is structured as follows: Section \ref{sec:modelling} gives a description of the \gls{AMB} system and its mathematical dynamical model. The proposed control design along with the stability analysis of the closed-loop system is detailed in Section \ref{sec:control}. Section \ref{sec:simulations} presents the simulation results and discusses the performance of the proposed method. Finally, concluding remarks and reflections on future work are presented in Section \ref{sec:conclusions}.

\section{System description} \label{sec:modelling}
	The system considered in this study comprises a vertical-axis rotor enclosed into two \gls{PMB} that stabilise the tilt angle to zero degrees, i.e. they passively guarantee the verticality of the axis \citep{andersen2013dynamics}. Two electromagnets, one on the top and one at the bottom of the rotor can be independently activated by the currents induced into their coils and accelerate the rotor mass along the vertical direction $z$ as shown in Fig. \ref{fig:AMB_system}. A common practice is to relate the currents of the two electromagnets $i_{up}$, $i_{down}$ through a common deviation $i$ from a known constant current value $i_0$, such that $i_{up} = i_0 + i$ and $i_{down} = i_0 - i$ \citep{maslen2009magnetic,chiba2005magnetic}. This achieves the reduction of the control inputs from two to one, i.e. the current deviation $i$. This approach is also adopted in this study. The equations of motion along the z-axis and the dynamics of the current read \citep{dagnaes2018magnetic}:
\begin{figure}
	\subfloat[\centering]{\includegraphics[width=0.45\columnwidth]{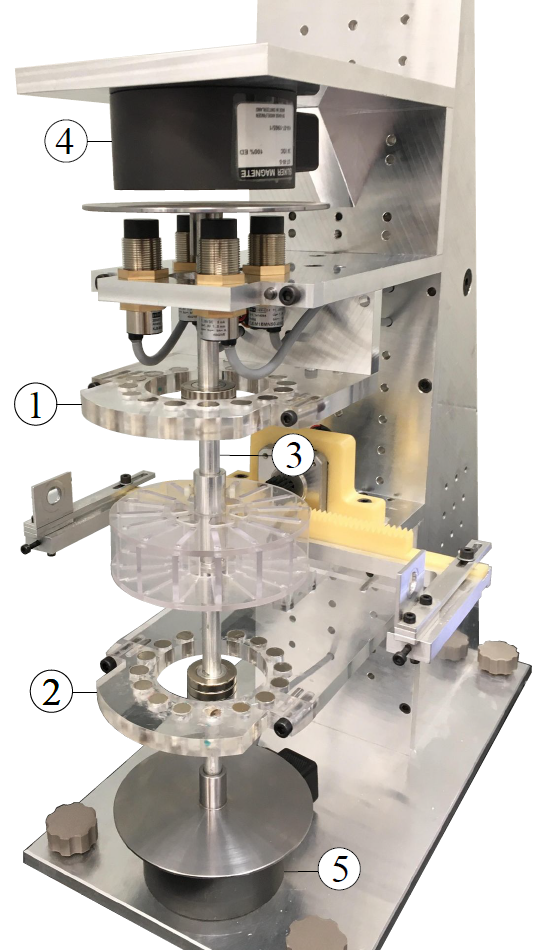} }
	\qquad
	\subfloat[\centering]{\includegraphics[width=0.45\columnwidth]{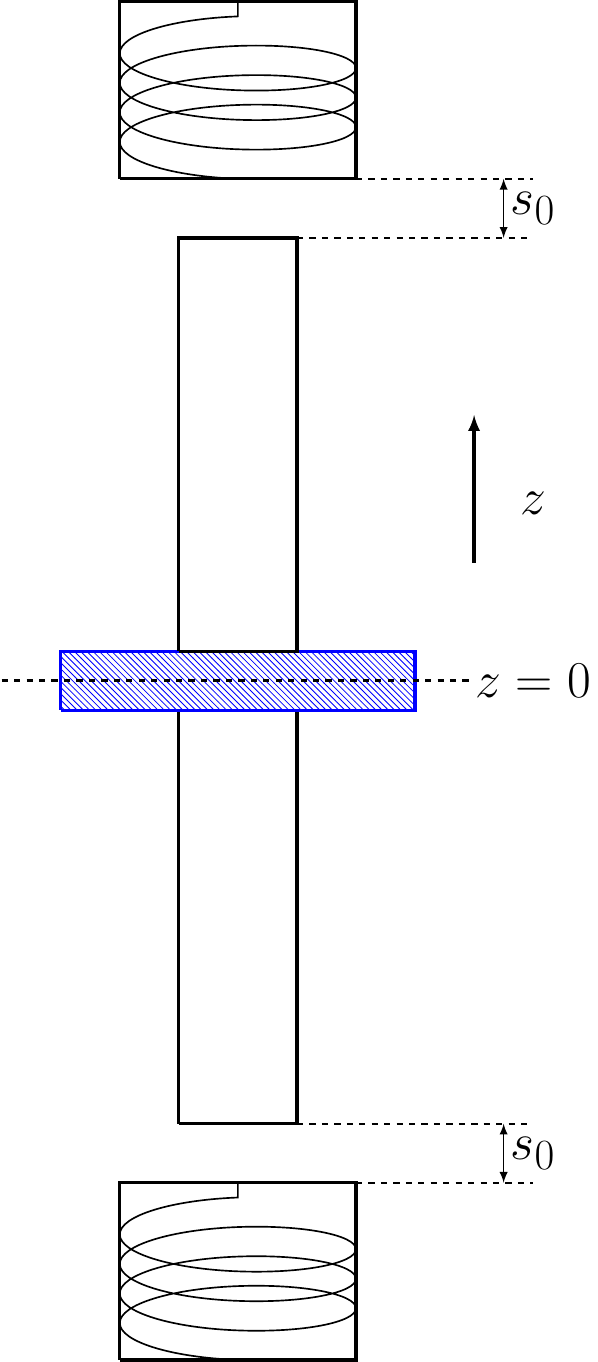} }
	\caption{(a) Typical rotating machine with active magnetic bearings: 1) Upper PMB, 2) Lower PMB, 3) Rotor, 4) Upper electromagnet and 5) Lower electromagnet. (b) Simplified diagram of the \gls{AMB} system.}
	\label{fig:AMB_system}
\end{figure}
\begin{align}
	\ddot{z} &= \frac{2 k_z}{m}z + \frac{\mu_0 n^2 A}{4m}\left[ \left( \frac{i_0 + i}{s_0 - z} \right)^2 - \left( \frac{i_0 - i}{s_0 + z} \right)^2 \right] - g + q_z \label{eq:z_dynamics}\\
	\frac{di}{dt} &= \frac{2 \dot{z}}{s_0 + z}i + \frac{2(s_0 + z)}{\mu_0 n^2 A}(u - Ri) + q_i \label{eq:z_dot_dynamics}
\end{align}
where $m$ is the rotor mass, $z$ is the vertical displacement of the rotor, $u$ is the voltage input, $s_0$ is the air gap between rotor and electromagnets at the equilibrium point, $\mu_0$ is vacuum permeability, $n$ is the number of windings in each coil, $R$ is its resistance, $A$ is its cross-sectional area, $k_z$ is the axial stiffness coefficient and $g$ is the gravitational acceleration. The unknown perturbations $q_z(t)$ and $q_i(t)$ are bounded with $\vert q_z(t) \vert \leq Q_z, \; \vert q_i(t) \vert \leq Q_i$, $\forall t \geq 0$ and account for the lumped model uncertainties and disturbances in the two subsystems. Table \ref{tab:par_tab} lists the values of the system parameters.
\begin{table}[bp]
	\begin{center}
		\caption{System parameter values}
		\label{tab:par_tab}
		\begin{tabular}{clc}
	\toprule
	\textbf{Symbol}	& \textbf{Description} & \textbf{Value}\T\B \\
	\specialrule{.2em}{.1em}{-1em}\\
	$m$ 	& Mass of rotor and axle	& $0.588 \; \si{\kilogram}$\T\B \\
	$k_z$ 	& Axial stiffness 			& $-754 \; \si{\newton\meter}$\T\B \\
	$\mu_0$ & Vacuum permeability		& $1.25\cdot 10^{-6} \; \si{\newton\per\ampere^2}$\T\B \\
	$n$ 	& Number of coil windings	& $1480$\T\B \\
	$A$ 	& Cross-sectional area 		& $0.121 \; \si{\meter^2}$\T\B \\
	$s_0$ 	& Air gap size 				& $5\cdot 10^{-3} \; \si{\meter}$\T\B \\
	$i_0$ 	& Bias current				& $0.25 \; \si{\ampere}$\T\B \\
	$R$ 	& Coil resistance			& $41.44 \; \si{\ohm}$\T\B \\
	$g$ 	& gravitational acceleration& $9.81 \; \si{\meter\per\sec^2}$\T\B \\\hline
\end{tabular}
	\end{center}
\end{table}

\section{Control design}\label{sec:control}
	\subsection{Architecture}
\begin{figure}[t]
	\begin{center}
		\includegraphics[width=0.95\columnwidth]{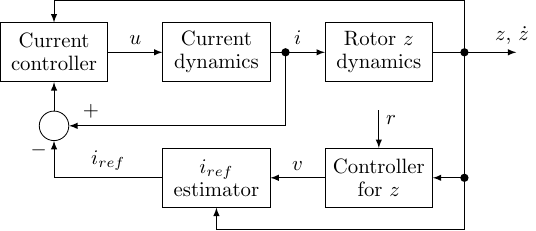}
		\caption{Control architecture for the \gls{AMB} system.}
		\label{fig:architecture}
	\end{center}
\end{figure}

	Consider the system in \eqref{eq:z_dynamics}, \eqref{eq:z_dot_dynamics} where all three states are available from measurements. The objective is to control the rotor axle vertical position such that the displacement $z$ is regulated at zero. The position and velocity of the rotor can be controlled by means of the current deviation $i$, which in turn can be controlled through the voltage $u$. Instead of employing backstepping strategies that often lead to complex designs, this study will pursue a modular architecture starting from defining the scaled magnetic force as virtual control input:
\begin{equation} \label{eq:v_definition}
	v(z,i) \triangleq \displaystyle \left( \frac{i_0 + i}{s_0 - z} \right)^2 - \left( \frac{i_0 - i}{s_0 + z} \right)^2, \; z\in(-s_0,s_0) \; .
\end{equation}
This essentially renders the mechanical subsystem an undamped mass-spring system. Next, an adaptive estimator will be deployed to invert the nonlinear mapping from the current deviation $i$ to the scaled magnetic force $v(z,i)$ such that the appropriate current reference signal is generated. Once the reference is obtained, the current dynamics will be regulated to track it such that the appropriate magnetic force, demanded by the controller for the mechanical system is generated and applied on the axle mass to position it at $z = 0$. \gls{SMC} will be used in both the mechanical and the electrical systems to ensure robust finite-time convergence of the controlled variable to the desired values. The overall control architecture is illustrated in Fig. \ref{fig:architecture}.
\tikzsetnextfilename{cascade}
	
	\subsection{Position control law}
	Given a smooth position reference signal $r(t)$ and the associated tracking error $e_z \triangleq z - r$, define the sliding variable $\sigma \triangleq \dot{e}_z + c e_z$, where $c > 0$. On the sliding manifold $\mathcal{S}: \sigma = 0$ the dynamics of the vertical displacement $z$ are governed by $\dot{e}_z = -c e_z$, i.e. $e_z^{*} = 0$ is a \gls{GES} equilibrium point, with rate of decay $c$. The dynamics of the sliding variable reads:
\begin{align}
	\dot{\sigma} &= \frac{2 k_z}{m}z + \frac{\mu_0 n^2 A}{4m}\left[ \left( \frac{i_0 + i}{s_0 - z} \right)^2 - \left( \frac{i_0 - i}{s_0 + z} \right)^2 \right] \nonumber\\
				 &- g + q_z + c\dot{e}_z - \ddot{r} \label{eq:sigma_dynamics}
\end{align}
Assigning the value
\begin{equation} \label{eq:z_control_law}
	v^{*} = \frac{4m}{\mu_0 n^2 A} \left[ -\frac{2k_z}{m}z - c(\dot{z} - \dot{r}) + \ddot{r} + g - k\text{sgn}(\sigma) \right]
\end{equation}
to the virtual control input defined in \eqref{eq:v_definition}, where $k > Q_z$ and the signum function is defined by
$$
\text{sgn}(x) \triangleq \begin{cases}
	\displaystyle \frac{x}{\vert x \vert} &\text{, for } x \neq 0\\
	\chi\in[ -1, 1 ] &\text{, for } x = 0
\end{cases}
$$
will drive the rotor axial dynamics onto the sliding manifold $\mathcal{S}$ in finite time \citep{slotine1991applied}. Calculating the appropriate value of the current directly from \eqref{eq:v_definition} can be problematic due to the possible existence of two values for $i$ (provided that the equation has real roots). Numerical ill-posedness is another potential challenge especially under the presence of sensor noise. To overcome these obstacles, an online estimator is designed that dynamically inverts the nonlinear mapping (see \citep{nicosia1994nonlinear} for more details) $v^{*} = v(z,i)$ with respect to $i$, providing thus a smooth estimate $i_{ref}$ of the appropriate current value. The design of this estimator is detailed in the next subsection.
	
	\subsection{Input mapping inversion}
	Let $i^{*}$ be an appropriate value of the current $i$ such that $v(z,i^{*}) = v^{*}$, with $v^{*}$ given in \eqref{eq:z_control_law}. Note that since $v^{*}$ is a state feedback control law, the value for $i$ is generally time-varying. Define the input error $\tilde{v} \triangleq v^{*} - v(z,i_{ref}) = v(z,i^{*}) - v(z,i_{ref})$ and $d(z,\dot{z},\dot{r},\ddot{r}) \triangleq \dot{v}^{*} - \frac{\partial v(z,i_{ref})}{\partial z}\dot{z}$.
\begin{assumption}\label{assum:delta_boundeness}
	The signal $d(z,\dot{z},r,\dot{r},\ddot{r})$ is bounded, i.e. $\exists\Delta_1 > 0$, such that $\vert d(z,\dot{z},r,\dot{r},\ddot{r}) \vert \leq \Delta_1, \; \forall t \geq 0$.
\end{assumption}
\begin{prop} \label{prop:adaptive_law}
	The adaptive law
	\begin{equation} \label{eq:adaptive_law}
		\frac{d i_{ref}}{dt} = \gamma \left( \frac{\partial v(z,i_{ref})}{\partial i_{ref}} \right)^{-1}\text{sgn}(\tilde{v})
	\end{equation}
	with $\gamma > 0$ such that
	\begin{equation} \label{eq:gamma_selection}
		\gamma > \Delta_1 \geq \left \vert \dot{v}^{*} - \frac{\partial v(z,i_{ref})}{\partial z}\dot{z} \right \vert
	\end{equation}
	stabilises the input error $\tilde{v}$ dynamics at the origin in finite time.
\end{prop}
\begin{pf}
	The dynamics of the input error reads:
	\begin{align*}
		\dot{\tilde{v}} &= \dot{v}^{*} - \dot{v}(z,i_{ref}) = \underbrace{\dot{v}^{*} - \frac{\partial v(z,i_{ref})}{\partial z}\dot{z}}_{d(z,\dot{z},\dot{r},\ddot{r})} - \frac{\partial v(z,i_{ref})}{\partial i_{ref}}\frac{d i_{ref}}{dt}
	\end{align*}
	Inserting the adaptive law \eqref{eq:adaptive_law} in the equation above gives
	\begin{equation}\label{eq:inversion_error_dynamics}
		\dot{\tilde{v}} = -\gamma\text{sgn}(\tilde{v}) + d(z,\dot{z},r,\dot{r},\ddot{r})
	\end{equation}
	Under assumption \ref{assum:delta_boundeness}, selecting $\gamma > \Delta_1$ ensures convergence of $\tilde{v}$ to the origin in finite time \citep{shtessel2014sliding}. $\blacksquare$
\end{pf}
\begin{remark}
	 The adaptative law in \eqref{eq:adaptive_law} is implementable only if the gradient $\frac{\partial v(z,i_{ref})}{\partial i_{ref}}$ is not vanishing, i.e. only if changes in $i_{ref}$ affect the input estimation error. This translates to the requirement that $\forall t \geq 0$ the points $(z(t),i_{ref}(t))$ do not belong to the zero set of $\frac{\partial v(z,i_{ref})}{\partial i_{ref}}$, which is the curve $z^2 + \frac{2s_0}{i_0}i_{ref}z + s_0^2 = 0$, i.e. $\forall t \geq 0$
	 \begin{equation} \label{cond:gradient_non_zero}
	 	(z,i_{ref})\notin \mathcal{C} = \left\lbrace z,i_{ref}\in \mathbb{R} \big | z^2 + \frac{2s_0}{i_0}i_{ref}z + s_0^2 = 0 \right\rbrace \; .
	 \end{equation}
 	This can be seen as a controllability condition for the input error dynamics with $\frac{di_{ref}}{dt}$ being the ``input" to the system.
\end{remark}
\vspace{5pt}
\begin{remark}
	The result of Proposition \ref{prop:adaptive_law} does not depend on the selection of the control law $v^{*}$ for the $z$ dynamics. In fact the signals $\dot{v}^{*}$, $\frac{\partial v(z,i_{ref})}{\partial z}\dot{z}$ can be computed at each time instant so long $v^{*}$ is differentiable almost everywhere.
\end{remark}
% To obtain an estimation of $\Delta_1$, the sgn$(x)$ function can be approximated by a continuous function, such $\frac{2}{\pi}\arctan(px), \; p \gg 1$.
	
	\subsection{Current control law}
	The task of the current controller is to ensure that the appropriate current deviation $i_{ref}$ is generated by the \gls{AMB} coils such that the demanded total magnetic force is applied to the rotor mass for its axial positioning. Let the current tracking error be denoted by $e = i - i_{ref}$. Its dynamics reads:
\begin{equation} \label{eq:current_dynamics}
	\dot{e} = \frac{di}{dt} - \frac{di_{ref}}{dt} = \frac{2 \dot{z}}{s_0 + z}i + \frac{2(s_0 + z)}{\mu_0 n^2 A}(u - Ri) + q_i - \frac{di_{ref}}{dt} \; .
\end{equation}
The control law
\begin{equation} \label{eq:i_control_law}
	u = \frac{\mu_0 n^2 A}{2(s_0 + z)}\left[ -\frac{2 \dot{z}}{s_0 + z}i + \frac{di_{ref}}{dt} - k_i\text{sgn}(e) \right] + Ri, \; k_i > Q_i
\end{equation}
gives the closed-loop dynamics $\dot{e} = -k_i\text{sgn}(e) + q_i$, which has a finite-time stable equilibrium point at the origin. Note that $\frac{di_{ref}}{dt}$ is known from \eqref{eq:adaptive_law} by design.
	
	\subsection{Stability analysis}
	The rotor sliding variable closed-loop dynamics can be written as
\begin{align}
	&\dot{\sigma} = \frac{2 k_z}{m}z - g + q_z - \ddot{r} + c\dot{e}_z + \frac{\mu_0 n^2 A}{4m}v(z,i) = \frac{2 k_z}{m}z - g \nonumber\\
				 &+ q_z - \ddot{r} + c\dot{e}_z + \frac{\mu_0 n^2 A}{4m} \left[ v^{*} - \tilde{v} - v(z,i_{ref}) + v(z,i) \right] \nonumber\\
				 &= -k\text{sgn}(\sigma) + q_z + \underbrace{\frac{\mu_0 n^2 A}{4m} \left[ v(z,i_{ref} + e) - \tilde{v} - v(z,i_{ref}) \right]}_{h(z,e,\tilde{v})} \nonumber\\
				 &= -k\text{sgn}(\sigma) + q_z + h(z,e,\tilde{v}) \; .
\end{align}
Together with the dynamics of the input estimation error and the current tracking error, they comprise a feedback interconnection. It is easy to see that when $e = \tilde{v} = 0$ the unperturbed $\sigma$-dynamics have a finite-time stable equilibrium at the origin. Inspired by the approach proposed in \citep{lor2008a}, this feedback can be also viewed as a \emph{cascaded interconnection} of the systems
\begin{align}
	(\Sigma_1) &: \dot{\sigma}  = -k\text{sgn}(\sigma) + q_z + h(z,e,\tilde{v})\\
	(\Sigma_2) &: \bm{\dot{\xi}} \triangleq \begin{bmatrix}
		\dot{\tilde{v}}\\
		\dot{e}
	\end{bmatrix} = \begin{bmatrix}
		-\gamma\text{sgn}(\tilde{v}) + d(z,\sigma - cz)\\
		-k_i\text{sgn}(e) + q_i
	\end{bmatrix}
\end{align}
where the solutions of $(\Sigma_2)$ depend on the \emph{parameter} $\sigma(t;t_0,\sigma_0)$.
\setcounter{thm}{0}
\begin{thm}
	Under the assumption that condition \eqref{cond:gradient_non_zero} holds with $k > Q_z, \; k_i > Q_i$ and $\gamma > \Delta_1$, the closed loop system $(\Sigma_1)-(\Sigma_2)$ has a \gls{UAS} equilibrium point at the origin.
\end{thm}
\begin{pf}
	The unperturbed system $(\Sigma_1)$ with $e = \tilde{v} = 0 \Rightarrow h(z,e,\tilde{v}) = 0$ has a finite-time stable equilibrium point at the origin. This implies the existence of a $\mathcal{C}^1$ positive definite and radially unbounded Lyapunov function $V \triangleq \sigma^2$, for which it holds $\dot{V} \leq -2\bar{k}\vert \sigma \vert$, where $\bar{k} = k - Q_z$. In order to prove that the origin is \gls{UAS}, it is enough to show that Assumptions 1,4,5,7 and the conditions of Theorem 2 from \citep{lor2008a} are satisfied. In the subsequent analysis, the notation introduced in \citep{lor2008a} is adopted to facilitate direct referencing to the original formulation of the assumptions and Theorem 2.
	
	The finite-time stability of the origin of the unperturbed $(\Sigma_1)$ satisfies Assumptions 1 and 5 that require \gls{UGAS} origin instead. Define $V_1(\sigma) \triangleq V(\sigma)$ and the class $\mathcal{K}_{\infty}$ functions $\alpha_1(\sigma) \triangleq \vert \sigma \vert^2 = V_1(\sigma)$, $\alpha_4(\sigma) \triangleq 2\vert \sigma \vert$ and $\alpha_4^{\prime}(\Vert \bm{\xi} \Vert) \triangleq \frac{k\mu_0 n^2 A}{4m}\sqrt{L_v^2 + 1}\Vert \bm{\xi} \Vert$, where $L_v$ is the Lipschitz constant of $v(z,i_{ref} + e)$ with respect to $e$. Moreover, evaluating the time derivative of $V_1$ along the trajectories of the perturbed system gives:
	\begin{align*}
		\dot{V}_1 &\leq -2\bar{k}\vert \sigma \vert + 2k\vert \sigma \vert \frac{\mu_0 n^2 A}{4m}\vert v(z,i_{ref} + e) - \tilde{v} - v(z,i_{ref}) \vert\\
				  &\leq 2k\vert \sigma \vert \frac{\mu_0 n^2 A}{4m}\left( L_v\vert e \vert + \vert \tilde{v} \vert \right) = 2k\vert \sigma \vert \frac{\mu_0 n^2 A}{4m} \begin{bmatrix}
				  		L_v & 1
				  \end{bmatrix} \begin{bmatrix}
				  		\vert e \vert\\
				  		\vert \tilde{v} \vert
			  	  \end{bmatrix}\\
		  	  	  &\leq 2k\vert \sigma \vert \frac{\mu_0 n^2 A}{4m}\sqrt{L_v^2 + 1}\Vert \bm{\xi} \Vert = \alpha_4(\sigma) \alpha_4^{\prime}(\Vert \bm{\xi} \Vert)
	\end{align*}
	Furthermore, if $V_{1,0}$ is a lower bound for $V_1$, then
	$$
		\int_{V_{1,0}}^{\infty} \frac{dw}{\alpha_4(\alpha_1^{-1}(w))} = \int_{V_{1,0}}^{\infty} \frac{dw}{2\sqrt{w}} = \left[ \sqrt{w} \right]_{V_{1,0}}^{\infty} = \infty
	$$
	which shows that Assumption 4 from \citep{lor2008a} is also satisfied. The finite-time stability of the origin of $(\Sigma_2)$ - note that this holds irrespectively of the dependence on the trajectories of $(\Sigma_1)$ - satisfies Assumption 7, which requires \gls{UGAS} for ($\Sigma_2$).
	
	The second and final condition of Theorem 2 in \citep{lor2008a} requires that there exist class $\mathcal{K}$ functions $\alpha_5(\sigma)$, $\alpha_5^{\prime}(\Vert \bm{\xi} \Vert)$ such that 
	$$
		\left \vert \frac{\partial V}{\partial \sigma}h(z,e,\tilde{v}) \right \vert \leq \alpha_5(\sigma)\alpha_5^{\prime}(\Vert \bm{\xi} \Vert)
	$$
	where $V$ was defined earlier. Since $V = V_1$, the foregoing inequality was shown to hold for $\alpha_5(\sigma) \triangleq \alpha_4(\sigma) = 2\vert \sigma \vert$ and $\alpha_5^{\prime}(\Vert \bm{\xi} \Vert) \triangleq \alpha_4^{\prime}(\Vert \bm{\xi} \Vert) = \frac{\mu_0 n^2 A}{4m}\sqrt{L_v^2 + 1}\Vert \bm{\xi} \Vert$. Moreover, it is required that for every positive upper bound $\rho$ of the solutions of $(\Sigma_2)$ $\exists \lambda_{\rho}, \eta_{\rho} > 0$ such that
	$$
		\forall t \geq 0 \; , \vert \sigma \vert \geq \eta_{\rho} \Rightarrow \alpha_5(\sigma) \leq \lambda_{\rho} W(\sigma)
	$$
	where $W$ is a positive semi-definite function. Indeed, selecting $W(\sigma) \triangleq 2\sigma^2$ one obtains:
	$$
		W(\sigma) = 2\sigma^2 \geq \eta_{\rho}2\vert \sigma \vert = \eta_{\rho}\alpha_5(\sigma) \Leftrightarrow \alpha_5(\sigma) \leq \lambda_{\rho}W(\sigma)
	$$
	with $\lambda_{\rho} \triangleq \frac{1}{\eta_{\rho}}$. With this, all conditions of Theorem 2 are satisfied and by Proposition 2 in  \citep{lor2008a}, the origin of $(\Sigma_1) - (\Sigma_2)$ is a \gls{UAS} equilibrium point. $\blacksquare$
\end{pf}
\begin{remark}
	Selecting $k > Q_z, \; k_i > Q_i$ suffices for ensuring finite-time stability of the origins of $\sigma$ and $e$ in the presence of unmodelled dynamics and perturbations in the system. Following the same line of argumentation, $k$ can be chosen larger than $\sup\limits_{t} \left\vert q_z(t) + h(z(t),e(t),\tilde{v}(t)) \right\vert$ to even dominate over the effects of the transients due to the feedback interconnection with $(\Sigma_2)$. Moreover, sgn$(x)$ can be approximated by a continuous function, such $\frac{2}{\pi}\arctan(px), \; p \gg 1$ to alleviate the effect of chattering.
\end{remark}
\begin{remark}
	The control laws \eqref{eq:z_control_law} and \eqref{eq:i_control_law} are only two of several possible options. In fact, provided that condition \eqref{cond:gradient_non_zero} holds, \gls{UAS} of the closed-loop system ($\Sigma_1$)-($\Sigma_2$) is ensured for any selection of $v^{*}$ and $u$ that render the origins of \eqref{eq:sigma_dynamics}, \eqref{eq:current_dynamics} \gls{UGAS}. It should also be noted that \gls{UGAS} of the origin of the cascaded system cannot be claimed since trajectories starting in $\mathcal{C}$ will automatically result in violation of condition \eqref{cond:gradient_non_zero}, which is essential for the finite time stability of the origin of \eqref{eq:inversion_error_dynamics}.
\end{remark}

\section{Simulation results} \label{sec:simulations}
	The efficacy and performance of the proposed control scheme were tested in two simulation scenarios, namely:
\begin{itemize}
	\item Regulation of $z$ at the origin, i.e. $r(t) \equiv 0$.
	\item Tracking of sinusoidal signal $r(t) = A_r\sin(2\pi f_r t)$.
\end{itemize}
\begin{table}[bp]
\begin{center}
	\caption{Simulation parameter values}
	\label{tab:spar_tab}
	\begin{tabular}{clc}
	\toprule
	\textbf{Symbol}	& \textbf{Description} & \textbf{Value}\T\B \\
	\specialrule{.2em}{.1em}{-1em}\\
	$A_r$ 	& Reference amplitude				& $0.0025 \; \si{\meter}$\T\B \\
	$f_r$ 	& Reference frequency 				& $2 \; \si{\hertz}$\T\B \\
	$c$ 	& Exponential decay rate			& $17$\T\B \\
	$k$ 	& \gls{SMC} gain for $z$			& $25$\T\B \\
	$\gamma$& Adaptation gain 					& $1000$\T\B \\
	$k_i$ 	& \gls{SMC} gain for $i$ 			& $152$\T\B \\
	$p$ 	& Signum approximation steepness factor	& $25$\T\B \\
	$Q_z$ 	& Disturbance amplitude				& $1$\T\B \\\hline
\end{tabular}
\end{center}
\end{table}
A pulse scaled magnetic force disturbance $q_z(t)$ of amplitude $Q_z$ is introduced in both cases to assess the robustness of the closed-loop system. It is assumed that the position sensor is inflicted with zero mean Gaussian noise with variance $10^{-7}$ m, which corresponds to approximately $20\%$ of the allowable axial displacement $s_0$ of the rotor. Table \ref{tab:spar_tab} lists all the simulation parameters.

\begin{figure}[t]
	\begin{center}
		\includegraphics[width=0.95\columnwidth]{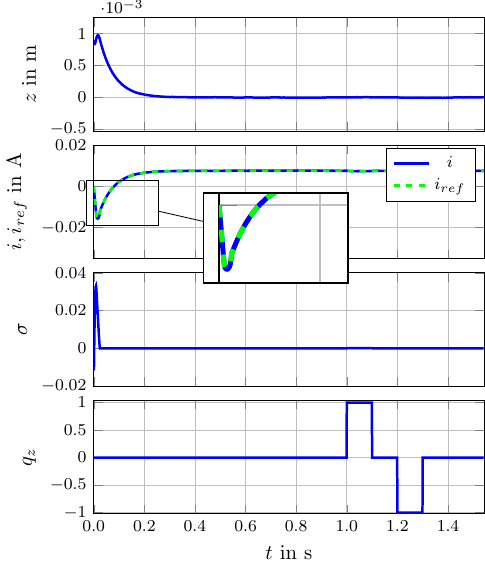}
		\caption{Robust regulation of axle position at $z = 0$. From top to bottom: Rotor axle deviation from zero, reference and actual current, sliding variable, disturbance.}\label{fig:signals_z}
	\end{center}
\end{figure}

\begin{figure}[t]
	\begin{center}
		\includegraphics[width=0.95\columnwidth]{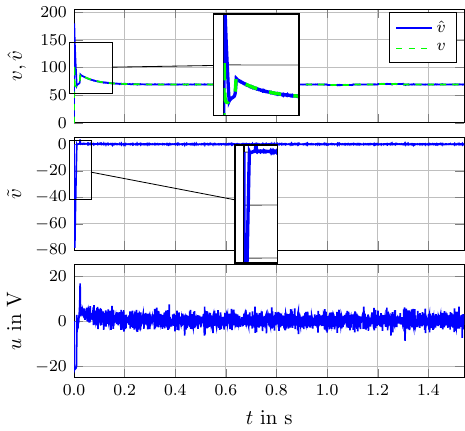}
		\caption{Robust regulation of axle position at $z = 0$. From top to bottom: required and generated scaled magnetic force, scaled magnetic force error, voltage input.}\label{fig:signals_u}
	\end{center}
\end{figure}

\begin{figure}[t]
	\begin{center}
		\includegraphics[width=0.95\columnwidth]{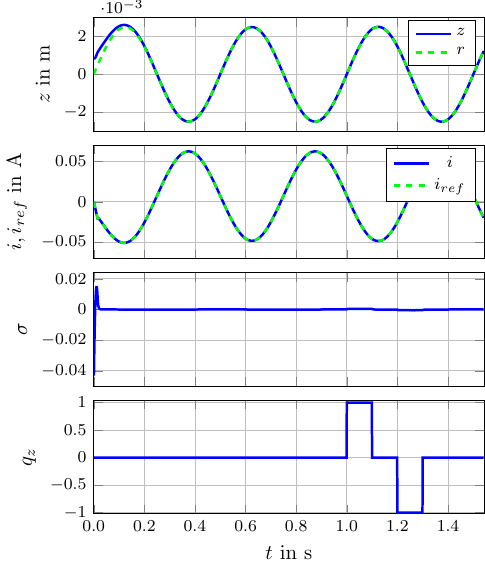}
		\caption{Robust tracking of axle position profile. From top to bottom: Rotor axle deviation from zero, reference and actual current, sliding variable, disturbance.}\label{fig:signals_z_track}
	\end{center}
\end{figure}

\begin{figure}[t]
	\begin{center}
		\includegraphics[width=0.95\columnwidth]{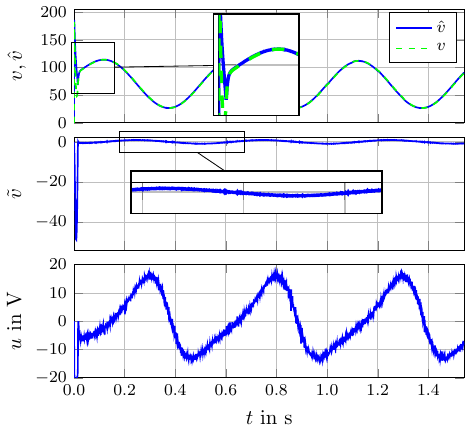}
		\caption{Robust tracking of axle position profile. From top to bottom: required and generated scaled magnetic force, scaled magnetic force error, voltage input.}\label{fig:signals_u_track}
	\end{center}
\end{figure}
Figure \ref{fig:signals_z} illustrates the response of the closed-loop system in the case of regulation at $z = 0$. Both the position displacement and the current bias converge to the reference signals, while the sliding surface is reaching a neighbourhood of the origin (due to noise) with radius $6\cdot 10^{-6} \; \si{\meter}$ in finite time. The closed-loop system is insensitive to the effect of the disturbance. The performance of the input mapping inversion scheme is shown in Fig. \ref{fig:signals_u} with the input estimation error reaching zero in less than 0.01 sec. Similar conclusions can be drawn by looking at Fig. \ref{fig:signals_z_track} and \ref{fig:signals_u_track}, which show the closed-loop response during the tracking case. The position displacement $z$ of the rotor is able to follow the sinusoidal reference signal with maximum absolute error $2.3\cdot 10^{-5} \; \si{\meter}$, while at the same time the disturbance is successfully rejected.

%%
%% \begin{thm} ... \end{thm}            % Theorem
%% \begin{lem} ... \end{lem}            % Lemma
%% \begin{claim} ... \end{claim}        % Claim
%% \begin{conj} ... \end{conj}          % Conjecture
%% \begin{cor} ... \end{cor}            % Corollary
%% \begin{fact} ... \end{fact}          % Fact
%% \begin{hypo} ... \end{hypo}          % Hypothesis
%% \begin{prop} ... \end{prop}          % Proposition
%% \begin{crit} ... \end{crit}          % Criterion

\section{Conclusions and future work} \label{sec:conclusions}
	A cascaded architecture for nonlinear control of an active magnetic bearings system was proposed in this study. The design employed two sliding mode control loops with an adaptive estimator for inversion of the nonlinear input mapping. The asymptotic stability of the closed-loop system origin was proven and verified in simulation. The results showed that the proposed control strategy achieves robust positioning of the rotor axle in the vertical axis as well as tracking of smooth position profiles and the same time it is able to reject bounded disturbances. Future extensions of this work will include experimental verification of the design and investigation of higher-order sliding mode controllers, such as the super twisting algorithm for robust control of \gls{AMB} systems.

\balance

\bibliography{mybibl}             % bib file to produce the bibliography

\begin{thebibliography}{25}
\providecommand{\natexlab}[1]{#1}
\providecommand{\url}[1]{\texttt{#1}}
\providecommand{\urlprefix}{URL }
\expandafter\ifx\csname urlstyle\endcsname\relax
  \providecommand{\doi}[1]{doi:\discretionary{}{}{}#1}\else
  \providecommand{\doi}{doi:\discretionary{}{}{}\begingroup
  \urlstyle{rm}\Url}\fi

\bibitem[{Andersen et~al.(2013)Andersen, Enemark, and
  Santos}]{andersen2013dynamics}
Andersen, S.B., Enemark, S., and Santos, I.F. (2013).
\newblock Dynamics and stability of rigid rotors levitated by passive
  cylinder-magnet bearings and driven/supported axially by pointwise contact
  clutch.
\newblock \emph{Journal of Sound and Vibration}, 332(25), 6637--6658.

\bibitem[{Chiba et~al.(2005)Chiba, Fukao, Ichikawa, Oshima, Takemoto, and
  Dorrell}]{chiba2005magnetic}
Chiba, A., Fukao, T., Ichikawa, O., Oshima, M., Takemoto, M., and Dorrell, D.G.
  (2005).
\newblock \emph{Magnetic bearings and bearingless drives}.
\newblock Elsevier.

\bibitem[{Dagn{\ae}s-Hansen(2018)}]{dagnaes2018magnetic}
Dagn{\ae}s-Hansen, N.A. (2018).
\newblock \emph{Magnetic Bearings for Offshore Flywheel Energy Storage
  Systems}.
\newblock Technical University of Denmark.

\bibitem[{Dong and You(2013)}]{dong2013a}
Dong, L. and You, S. (2013).
\newblock Adaptive back-stepping control of active magnetic bearings.
\newblock \emph{Ieee International Conference on Control and Automation, Icca},
  452--457.
\newblock \doi{10.1109/ICCA.2013.6564856}.

\bibitem[{Fittro and Knospe(2002)}]{fittro2002a}
Fittro, R.L. and Knospe, C.R. (2002).
\newblock The $\mu$ approach to control of active magnetic bearings.
\newblock \emph{Journal of Engineering for Gas Turbines and Power}, 124(3),
  566--570.
\newblock \doi{10.1115/1.1417484}.

\bibitem[{Gibson and Buckner(2002)}]{gibson2002a}
Gibson, N.S. and Buckner, G.D. (2002).
\newblock Real-time adaptive control of active magnetic bearings using linear
  parameter varying models.
\newblock \emph{Conference Proceedings - Ieee Southeastcon}, 268--272.

\bibitem[{Huynh and Hoang(2016)}]{huynh2016a}
Huynh, V.V. and Hoang, B.D. (2016).
\newblock Second order sliding mode control design for active magnetic bearing
  system.
\newblock \emph{Lecture Notes in Electrical Engineering}, 371, 519--529.
\newblock \doi{10.1007/978-3-319-27247-4-44}.

\bibitem[{Jang et~al.(2005)Jang, Chen, and Tsao}]{jang2005a}
Jang, M.J., Chen, C.L., and Tsao, Y.M. (2005).
\newblock Sliding mode control for active magnetic bearing system with flexible
  rotor.
\newblock \emph{Journal of the Franklin Institute}, 342(4), 401--419.
\newblock \doi{10.1016/j.jfranklin.2005.01.006}.

\bibitem[{Junfeng et~al.(2011)Junfeng, Kun, and Kai}]{junfeng2011a}
Junfeng, C., Kun, L., and Kai, X. (2011).
\newblock ${H}_{\infty}$ control of active magnetic bearings using closed loop
  identification model.
\newblock \emph{2011 Ieee International Conference on Mechatronics and
  Automation, Icma 2011}, 349--353.
\newblock \doi{10.1109/ICMA.2011.5985682}.

\bibitem[{Kandil et~al.(2018)Kandil, Dubois, Bakay, and Trovão}]{kandil2018a}
Kandil, M.S., Dubois, M.R., Bakay, L.S., and Trovão, J.P.F. (2018).
\newblock Application of second-order sliding-mode concepts to active magnetic
  bearings.
\newblock \emph{Ieee Transactions on Industrial Electronics}, 65(1), 7962229.
\newblock \doi{10.1109/TIE.2017.2721879}.

\bibitem[{Kang et~al.(2010)Kang, Lyou, and Lee}]{kang2010a}
Kang, M.S., Lyou, J., and Lee, J.K. (2010).
\newblock Sliding mode control for an active magnetic bearing system subject to
  base motion.
\newblock \emph{Mechatronics}, 20(1), 171--178.
\newblock \doi{10.1016/j.mechatronics.2009.09.010}.

\bibitem[{Lee et~al.(2013)Lee, Lee, Shin, Kim, and Chung}]{lee2013a}
Lee, Y., Lee, S.H., Shin, D., Kim, W., and Chung, C.C. (2013).
\newblock Position control of active magnetic bearings using linear parameter
  varying synthesis.
\newblock \emph{International Conference on Control, Automation and Systems},
  5--10.
\newblock \doi{10.1109/ICCAS.2013.6703854}.

\bibitem[{Long et~al.(1996)Long, Carroll, and Mukundan}]{long1996a}
Long, M.L., Carroll, J.J., and Mukundan, R. (1996).
\newblock Adaptive control of active magnetic bearings under unknown static
  load change and unbalance.
\newblock \emph{Ieee Conference on Control Applications - Proceedings},
  876--881.
\newblock \doi{10.1109/CCA.1996.558982}.

\bibitem[{Loría(2008)}]{lor2008a}
Loría, A. (2008).
\newblock From feedback to cascade-interconnected systems: Breaking the loop.
\newblock \emph{Proceedings of the IEEE Conference on Decision and Control},
  4109--4114.
\newblock \doi{10.1109/CDC.2008.4738647}.

\bibitem[{Maslen and Schweitzer(2009)}]{maslen2009magnetic}
Maslen, E.H. and Schweitzer, G. (2009).
\newblock \emph{Magnetic bearings: theory, design, and application to rotating
  machinery}.
\newblock Springer.

\bibitem[{Motee and de~Queiroz(2002)}]{motee2002a}
Motee, N. and de~Queiroz, M. (2002).
\newblock Control of active magnetic bearings with a smart bias.
\newblock \emph{Proceedings of the 41st Ieee Conference on Decision and
  Control, Vols 1-4}, 1, 860--865.
\newblock \doi{10.1109/CDC.2002.1184615}.

\bibitem[{Mouille and Lottin(1992)}]{mouille1992a}
Mouille, P. and Lottin, J. (1992).
\newblock Digital multivariable control of active magnetic bearings.
\newblock In \emph{System Structure and Control 1992}, 376--379. Elsevier.

\bibitem[{Nicosia et~al.(1994)Nicosia, Tornamb{\'e}, and
  Valigi}]{nicosia1994nonlinear}
Nicosia, S., Tornamb{\'e}, A., and Valigi, P. (1994).
\newblock Nonlinear map inversion via state observers.
\newblock \emph{Circuits, Systems and Signal Processing}, 13(5), 571--589.

\bibitem[{Saha et~al.(2020)Saha, Amrr, and Nabi}]{saha2020a}
Saha, S., Amrr, S.M., and Nabi, M. (2020).
\newblock Adaptive second order sliding mode control for the regulation of
  active magnetic bearing.
\newblock \emph{Ifac-papersonline}, 53(1), 1--6.
\newblock \doi{10.1016/j.ifacol.2020.06.001}.

\bibitem[{Schroder et~al.(1997)Schroder, Chipperfield, Fleming, and
  Grum}]{schroder1997a}
Schroder, P., Chipperfield, A.J., Fleming, P.J., and Grum, N. (1997).
\newblock Robust multivariable control of active magnetic bearings.
\newblock \emph{Ecc 1997 - European Control Conference}, 3537--3542.
\newblock \doi{10.23919/ecc.1997.7082662}.

\bibitem[{Shtessel et~al.(2014)Shtessel, Edwards, Fridman, Levant
  et~al.}]{shtessel2014sliding}
Shtessel, Y., Edwards, C., Fridman, L., Levant, A., et~al. (2014).
\newblock \emph{Sliding mode control and observation}, volume~10.
\newblock Springer.

\bibitem[{Slotine et~al.(1991)Slotine, Li et~al.}]{slotine1991applied}
Slotine, J.J.E., Li, W., et~al. (1991).
\newblock \emph{Applied nonlinear control}, volume 199.
\newblock Prentice hall Englewood Cliffs, NJ.

\bibitem[{Song and Mukherjee(1996)}]{song1996a}
Song, G. and Mukherjee, R. (1996).
\newblock Integrated adaptive robust control of active magnetic bearings.
\newblock \emph{Information Intelligence and Systems, Vols 1-4}, 3, 1784--1790.
\newblock \doi{10.1109/ICSMC.1996.565378}.

\bibitem[{Wang and Zhu(2016)}]{wang2016a}
Wang, Z. and Zhu, C. (2016).
\newblock Active control of active magnetic bearings for maglev flywheel rotor
  system based on sliding mode control.
\newblock \emph{2016 Ieee Vehicle Power and Propulsion Conference, Vppc 2016 -
  Proceedings}, 7791603.
\newblock \doi{10.1109/VPPC.2016.7791603}.

\bibitem[{Xu et~al.(2022)Xu, Zhou, and Xu}]{xu2022a}
Xu, B., Zhou, J., and Xu, L. (2022).
\newblock Adaptive backstepping control of active magnetic bearings with slice
  rotor.
\newblock \emph{Journal of Vibration Engineering and Technologies}, 10(2),
  795--808.
\newblock \doi{10.1007/s42417-021-00410-x}.

\end{thebibliography}
                                                     % with bibtex (preferred)

\end{document}